\documentclass[twocolumn,prb,multicol,amsmath,amssymb]{revtex4-2}
\usepackage[dvips]{graphicx}
\usepackage{graphicx}
\usepackage{dcolumn}
\usepackage{bm}
\usepackage{graphics}
\usepackage{epsfig,color}
\usepackage[normalem]{ulem} 
\usepackage{soul}

\newcommand{\be}{\begin{equation}}
\newcommand{\ee}{\end{equation}}
\newcommand{\bea}{\begin{eqnarray}}
\newcommand{\eea}{\end{eqnarray}}

\newcommand{\up}{\uparrow}
\newcommand{\down}{\downarrow}

\begin{document}
\title {Concurrence distribution in excited states of the 1D spin-1/2 transverse field XY model: two different regions }
\author{S. Mahdavifar$^{1}$}
\email[]{smahdavifar@gmail.com}
\author{Z. Balador$^{1}$}
\author{M. R. Soltani$^{2}$}
\affiliation{$^{1}$Department of Physics, University of Guilan, 41335-1914, Rasht, Iran}
\affiliation{$^{2}$Department of Physics, Yadegar-e-Imam Khomeini (RAH), Shahr-e-Rey Branch, Islamic Azad University, Tehran, Iran}

\begin{abstract}

We investigate the variation of concurrence in a spin-1/2 transverse field XY chain system in an excited state. Initially, we precisely solve the eigenvalue problem of the system Hamiltonian using the fermionization technique. Subsequently, we calculate the concurrence between nearest-neighbor pairs of spins in all excited states with higher energy than the ground state. Below the factorized  field, denoted as $h_f=\sqrt{J^2-(J \delta)^2}$, we find no pairwise entanglement between nearest neighbors in excited states. At the factorized field, corresponding to a factorized state, we observe weak concurrence in very low energy states. Beyond $h_f$, the concurrence strengthens, entangling all excited states. The density of entangled states peaks at the center of the excited spectrum. Additionally, the distribution of concurrence reveals that the midpoint of the non-zero concurrence range harbors the most entangled excited states.

\end{abstract}
\maketitle

\section{Introduction}\label{sec1} 

Quantum entanglement [{\color{blue}\onlinecite{E1,E2,E3,E4,E5}}] is a remarkable prediction of modern quantum mechanics. It occurs when two or more particles are so strongly correlated that their quantum states cannot be described independently, even when they are far apart. Entanglement describes a quantum property of a non-separable superposition state involving two or more quantum systems. Furthermore, it may reveal new connections between other areas of physics, such as condensed matter and statistical mechanics. Entanglement can lead to the recognition of exotic quantum phases, such as spin liquids [{\color{blue}\onlinecite{6,7,7-0}}], topological [{\color{blue}\onlinecite{8,9,9b}}], and many-body localized systems [{\color{blue}\onlinecite{10}}]. The exploration of entanglement in quantum many-body systems can be facilitated through experiments like full-state tomography [{\color{blue}\onlinecite{13,14}}]  and Renyi entropy measurements in ultra-cold atoms [{\color{blue}\onlinecite{15,16}}]. Recently, an experimental scheme based on inelastic neutron scattering was developed to detect and quantify entanglement in the solid-state. As an example, $\rm{Cs_2CoCl_4}$, a quasi one-dimensional spin-1/2 XXZ model with a transverse field, was used [{\color{blue}\onlinecite{22b}}].

Quantifying the entanglement of quantum systems with many parts, such as a system involving multiple qubits, is a key theoretical challenge in quantum entanglement theory. One common method to measure this entanglement is through the use of concurrence, which applies to both pure and mixed states of two qubits [{\color{blue}\onlinecite{E6}}]. The concurrence of a state ranges from zero for separable states to one for maximally entangled states.

For two arbitrary spins at position $i$ and $j$, the two-site
reduced density matrix generally takes the form  [{\color{blue}\onlinecite{E6-00}}],

\begin{eqnarray}\label{eq01}
\rho_{i,j}&=&\frac{1}{4}+\sum\limits_{\alpha} (\langle S_{i}^{\alpha} \rangle S_{i}^{\alpha}+\langle S_{j}^{\alpha} \rangle S_{j}^{\alpha})\\ \nonumber
&+&\sum\limits_{\alpha, \beta} \langle S_{i}^{\alpha}  S_{j}^{\beta} \rangle S_{i}^{\alpha}  S_{j}^{\beta},
\end{eqnarray}
where $\alpha, \beta=x, y, z$. The concurrence between two spin-1/2 particle at sites $i$ and $j$ can be obtained  from
the corresponding reduced density matrix $\rho_{ij}$.  The reduced density matrix in the standard basis ($|\uparrow\uparrow\rangle, |\uparrow\downarrow\rangle, |\downarrow\uparrow\rangle, |\downarrow\downarrow\rangle $) is expressed as

\begin{equation}\label{eq6}
{\rho _{i,j}} = \left( {\begin{array}{*{20}{c}}
{{\langle p_{i}^{\up} p_{j}^{\up} \rangle}}&  {{\langle p_{i}^{\up} S_j^{-} \rangle}}  &{{\langle S_i^{-} p_{j}^{\up}  \rangle}}&{{\langle S_i^{-} S_j^{-}  \rangle}}\\
{{\langle p_{i}^{\up} S_j^{+} \rangle}}&{{\langle p_{i}^{\up} p_{j}^{\down} \rangle}}&{{\langle S_i^{-} S_j^{+} \rangle}}&{{\langle S_i^{-} p_{j}^{\down}  \rangle}}\\
{{\langle S_i^{+} p_{j}^{\up}  \rangle}}&{{\langle S_i^{+} S_j^{-}  \rangle}}&{{\langle p_{i}^{\down} p_{j}^{\up} \rangle}}&{{\langle p_{i}^{\down} S_j^{-} \rangle}}\\
{{\langle S_i^{+} S_j^{+}  \rangle}}&{{\langle S_i^{+} p_{j}^{\down}  \rangle}}&{{\langle p_{i}^{\down} S_j^{+}   \rangle}}&{{\langle p_{i}^{\down} p_{j}^{\down} \rangle}}
\end{array}} \right),
\end{equation}
where the brackets symbolize the physical state average and $p^{\up}=\frac{1}{2}+S^{z}$,  $p^{\down}=\frac{1}{2}-S^{z}$, $S^{\pm}=S^{x}\pm i S^{y}$. The concurrence between two spins is given through $C=\max (0,\lambda_1-\lambda_2-\lambda_3-\lambda_4$, where $\lambda_{i}$ is the square root of the eigenvalue of $R=\rho _{i,j} \tilde{\rho}_{i,j}$ and $\tilde{\rho}_{i,j}=(\sigma_i^{y} \otimes \sigma_j^{y}) \rho _{i,j}^{\star} (\sigma_i^{y} \otimes \sigma_j^{y}) $. 
Considering to the symmetry of the Hamitlonian most of the off-diagonal elements of the reduced density matrix $\rho _{i,j}$, will be zero. First, the translation invariance require that the density matrix satisfies $\rho _{i,j}=\rho _{i,i+r}$ for any position $i$. Then,  the 1D spin-1/2 transverse field XY model (Eq.~\ref{eq1}) is a $Z_2$-symmetric model which means that it is invariance under $\pi$-rotation around $z$ direction. This also implies that the density matrix commutes with the operator $S_{i}^{z}  S_{j}^{z}$. Following these symmetry properties, the density matrix must be symmetrical and only some elements of the reduced density matrix becomes non-zero [{\color{blue}\onlinecite{E6-1,E6-2}}],

\begin{equation}\label{eq6}
{\rho _{i,j}} = \left( {\begin{array}{*{20}{c}}
{{X_{i,j}^ +}}&0&0&{{F_{i,j}^*}}\\
0&{{Y_{i,j}^ +}}&{{Z_{i,j}^*}}&0\\
0&{{Z_{i,j}}}&{{Y_{i,j}^ -}}&0\\
{{F_{i,j}}}&0&0&{{X_{i,j}^ -}}
\end{array}} \right).
\end{equation}
Finally the concurrence is given by the following expression
\begin{eqnarray}
C&=& \max \{0, C_1,C_2 \}, \rangle,\nonumber\\
C_1&=& 2 (|Z_{i,j}|-\sqrt{X_{i,j}^{+} X_{i,j}^{-}}),\nonumber\\
C_2&=& 2 (|F_{i,j}|-\sqrt{Y_{i,j}^{+} Y_{i,j}^{-}}).
\label{Concurr}
\end{eqnarray}

One-dimensional quantum spin systems exhibit numerous non-classical properties related to spin entanglement. The Heisenberg model, along with its special cases like the Ising, XY, and XXZ models, describes the magnetic behavior of these systems. Quantum dots can be modeled using systems with XY interaction [{\color{blue}\onlinecite{E7,E8}}], leading to extensive studies exploring the key characteristics of this model. An additional advantage is that the eigenvalues of this model can be precisely determined through the Jordan-Wigner transformation.

The spin-1/2 transverse field (TF) XY chain model is a widely studied topic in physics. Its ground state exhibits two phases: an antiferromagnetic phase with a broken $Z2$ symmetry in the infinite system size limit, and a paramagnetic phase. These phases are separated by a quantum critical point at a specific value of the TF, denoted as $h=h_c$ [{\color{blue}\onlinecite{E9,E10,E11}}]. The mentioned quantum phase transition has been characterized by studying the concurrence [{\color{blue}\onlinecite{E12,E13,E14,E15,E16,E17}}]. The concurrence, serving as a reliable criterion for entanglement measurement, is maximal close to the critical field, and its derivatives signal the presence of a quantum phase transition at the critical points

 Entanglement in excited states of spin chains is a fascinating topic that reveals various aspects of quantum dynamics, critical phenomena, and quantum information [{\color{blue}\onlinecite{E17-01, E17-02,E17-03,E17-04,E17-05}}].  The entanglement entropy measures the quantum correlations between different parts of the system. For the 1D spin-1/2 TF XY model, some excited states have extensive entanglement entropy, unlike the ground state with logarithmic entanglement entropy [{\color{blue}\onlinecite{E17-01}}]. For one-dimensional spin chains described by conformal field theory, the entanglement entropy of excited states follows a universal law that depends on the scaling dimension and the central charge of the theory [{\color{blue}\onlinecite{E17-02}}]. For the spin-1/2 Heisenberg chain with antiferromagnetic interactions, the entanglement entropy of excited states is reduced by the bound states of particles [{\color{blue}\onlinecite{E17-03}}]. 
 
 Here, we focus on the concurrence distribution in excited states which is a topic that examines how the concurrence of a quantum system varies when it is in an excited state, i.e., a state with higher energy than the ground state.  The concurrence distribution can have useful applications for manipulating or measuring the quantum properties of excited states, such as in quantum metrology, quantum information, and quantum computation. For example, one can use the concurrence pattern to find the best states for improving the accuracy of quantum measurements, such as in interferometry or spectroscopy [{\color{blue}\onlinecite{E17-1}}]. One can also use the concurrence pattern to design and implement quantum algorithms that use excited states as resources, such as in quantum phase estimation or quantum simulation [{\color{blue}\onlinecite{E17-2}}]. It can be also applied to understand and characterize the dynamics and transitions of quantum systems in their excited states, such as in quantum chaos or quantum phase transitions [{\color{blue}\onlinecite{E17-3}}].

We utilize the 1D spin-1/2 TF XY model and apply the fermionization technique to diagonalize the system's Hamiltonian, extracting its eigenvalues and eigenvectors. The ground state corresponds to the vacuum state of the Bogoliubov fermion number operator. Excited states are then computed within subspaces defined by different values of this number operator.
Our analysis reveals two distinct regions, separated by the factorized point. In the first region, where $h < h_f$ , nearest neighbor pairs of spins exhibit no entanglement in excited states. Conversely, in the second region, they are entangled across all excited states. Notably, the density of entangled states peaks at the midpoint of the excited spectrum, and the most entangled excited states are concentrated in the middle of the non-zero range of the concurrence.

The paper is organized as follows. In the next section, we introduce the model and employ the fermionization approach to derive the system's spectrum. In Section III, we present our findings regarding the concurrence between nearest-neighbor pairs of spins in all excited states. Finally, in Section IV, we provide our conclusions and a summary of the results


\section{The model}

The Hamiltonian of the 1D spin-1/2 TF XY  model is defined as 
\begin{eqnarray}\label{eq1}
{\cal H} &=&J\sum\limits_{n = 1}^N {\left[ {(1 + \delta )S _n^x S _{n + 1}^x + (1 - \delta ) S _n^y S _{n + 1}^y} \right]} \nonumber \\
&-&h\sum\limits_{n = 1}^N {S _n^z}~,
\end{eqnarray}
where $S _n$ is the spin operator on the $n$-th site. $J>0$ denotes the antiferromagnetic exchange coupling.  $0 \leq \delta \leq 1$ and $h$ are the anisotropy parameter and the homogeneous TF, respectively.  $N$ is the system size (or number of spins) and we consider the  periodic boundary condition $S _{n+N}^\mu=S _n^\mu $ ($\mu=x,y,z$). $J=1$ is considered without losing generality.  

A notable feature occurs on the circle defined by $h_{f}^{2}+(J\delta)^2 = J^2$, referred to as the factorized point. At this point, the ground state wave function factorizes into a product of single-spin states [{\color{blue}\onlinecite{E18,E19}}]. The factorized point introduces distinct regimes in the model's phase diagram, particularly in terms of the revivals observed in the Loschmidt echo [{\color{blue}\onlinecite{E19-1}}].  The Loschmidt echo serves as a metric for gauging how susceptible a quantum system is to minor changes or imperfections. In an ideal scenario of perfect isolation and reversibility, the time-reversal procedure should seamlessly restore the system to its initial state. However, when subjected to sources of decoherence, such as experimental errors or environmental interactions, the time-reversal procedure fails to fully recover the initial state. The Loschmidt echo quantifies this deviation by comparing the final state with the initial state.


The ground state phase diagram also splits into two regions: $h <h_f$ where there is no spin squeezing, and $h >h_f$ where there is spin squeezing. The boundary between the two regions, $h = h_f$, supports spin coherence [{\color{blue}\onlinecite{E19-2}}].  Spin squeezing is a quantum process that decreases the uncertainty of one of the angular momentum components in a group of particles with a spin. The resulting quantum states obtained are referred to as  spin squeezed states.

The Hamiltonian can be diagonalized [{\color{blue}\onlinecite{E19-3,E17-02}}].   First, applying the Jordan-Wigner transformation, 
\begin{eqnarray}
S^{+}_{n}&=&a_{n}^{\dag}e^{i\pi\sum^{n-1}_{m=1}a^{\dag}_{m}a_{m}},\nonumber\\
S^{-}_{n}&=&e^{-i\pi\sum^{n-1}_{m=1}a^{\dag}_{m}a_{m}}a_{n},\nonumber\\
S^{z}_{n}&=&a_{n}^{\dag}a_{n}-\frac{1}{2},
\label{eq17}
\end{eqnarray}
where, $a_n^\dagger$ and $a_n$ are the fermionic operators, the fermionized form of the Hamiltonian is obtained as
\begin{eqnarray}\label{eq0-2}
{\cal H} &=& \sum\limits_{n = 1}^{N-1} \left[ \frac{1}{2} a_n^\dag a_{n + 1}+\frac{\delta}{2} a_n^\dag a_{n + 1}^\dag  + h.c. \right] \nonumber\\
& -& h\sum\limits_{n = 1}^N a_n^\dag {a_n}\nonumber \\
&-&\frac{\mu_p}{2} \left[ a_N^\dag a_{1}+ \frac{\delta}{2} a_N^\dag a_{1}^\dag +h.c.\right].
\end{eqnarray}
While the boundary terms in the Hamiltonian are typically negligible in the thermodynamic limit, where the number of particles and the system size are very large, they play a crucial role in determining the symmetry of the system. These terms establish the boundary conditions for the fermions, which are created or destroyed in pairs. As a consequence, the total number of fermions is either even or odd. This parity property is encapsulated by the parity operator $\mu_p  =\Pi^{N}_{n=1} S_n^{z}$, which commutes with the Hamiltonian, i.e., $[\mu_p,{\cal H}]=0$. This implies that the system can be categorized into two sectors with different parity, denoted by $\mu_p = \pm 1.$ The positive parity corresponds to the even sector, characterized by fermions with anti-periodic boundary conditions ($a_{n + N}=-a_{n }$). Conversely, the negative parity corresponds to the odd sector, where fermions exhibit periodic boundary conditions ($a_{n + N}=a_{n }$). With these definitions, the Hamiltonian can be written as

\begin{eqnarray}\label{eq2}
{\cal H} &=& \sum\limits_{n = 1}^{N} \left[ \frac{1}{2} a_n^\dag a_{n + 1}+\frac{\delta}{2} a_n^\dag a_{n + 1}^\dag  + h.c. \right] \nonumber\\
& -& h\sum\limits_{n = 1}^N a_n^\dag {a_n}.
\end{eqnarray}
Next, implementing  a Fourier transformation ${a_n} =  \sum_k e^{ - ikn} {a_k}$, and then applying a Bogoliuobov transformation 
${a_k} = \cos ({\theta _k}) {\beta _k} + i\sin ({\theta _k}) \beta _{ - k}^{\dag}$, yields the diagonalized  Hamiltonian  
\begin{equation}\label{eq3}
{\cal H} = \sum\limits_k {{\varepsilon }_k\left( {\beta _k^\dag {\beta _k} - \frac{1}{2}} \right)},
\end{equation}
with energy spectrum given by
\begin{equation} \label{spectrum}
\varepsilon_k = \sqrt{{\cal A}_k ^2+{\cal C}_k ^2},
\end{equation}
where
${\cal A}_k= \cos(k)-h$  and ${\cal C}_k=  -\delta \sin(k)$
are associated with the Bogoliubov angle $\theta_k$ by $\tan (2{\theta _k}) =- {\cal C}_k/{\cal A}_k$. It should be noted that the summation in Eq. (\ref{eq3}) runs over $k=2\pi m/N$, with $m=0,\pm 1,...,\pm \frac{1}{2}(N-1) \ [m= 0, \pm 1,..., \pm (\frac{1}{2}N-1), \frac{1}{2}N]$ for $N$ odd [$N$ even] (with periodic boundary conditions imposed on the Jordan-Wigner fermions).

The Hamiltonian and the total Bogoliubov fermion number operator, $\hat{N}_{B}=\sum\limits_k  \beta_k^\dag \beta_k $, commute with each other. The number operator $\hat{N}_{B}$ has eigenvalues $N_B={0,1,2,....,N}$, corresponding to the number of fermions in the system. The lowest energy state is the vacuum state with $N_B=0$. For each value of $N_B=m$, there are $\frac{N!}{m! (N-m)!}$  possible ways to arrange the fermions in the excited states. These states form energy bands that are indexed by $N_B=m$.


\section{Results}

We use concurrence, a measure of entanglement, to assess the degree of entanglement between two spins at sites $i$ and $j$. Focusing exclusively on spins that are adjacent (i.e., $j = i+1$), we calculate the concurrence using the reduced density matrix. This matrix describes the two-point correlation functions in terms of fermion operators, providing a comprehensive view of the entanglement between the pairs of spins

\begin{eqnarray}\label{eq7}
X^{+}&=& f_0^2-|f_1|^2+|f_2|^2,\nonumber\\
X^{-}&=& 1-2 f_0+X_{n,n+1}^{+}, \nonumber\\
Y^ {+}  &=& Y_{n,n + 1}^ {-}=f_0-X_{n,n+1}^{+}, \nonumber\\
Z&=&f_1, \nonumber\\
F &=&f_2, 
\end{eqnarray}
where
\begin{eqnarray}\label{eq8}
f_0&=&\frac{1}{N} \sum_{k} [ \cos(2{\theta _k})~\langle \beta _{k}^{\dag} \beta _{k} \rangle+\sin^2({\theta _k}) ], \nonumber \\
f_1&=&\frac{1}{N} \sum_{k} \cos(k)~[ \cos(2{\theta _k})~\langle \beta _{k}^{\dag} \beta _{k} \rangle+\sin^2({\theta _k}) ] \nonumber \\
&-&\frac{i}{N} \sum_{k} \sin(k)~\langle \beta _{k}^{\dag} \beta _{k} \rangle, \nonumber \\
f_2&=&\frac{1}{N} \sum_{k} [\sin(k)\sin(2 {\theta _k})]~ [-\frac{1}{2}+\langle \beta _{k}^{\dag} \beta _{k} \rangle ]. \nonumber \\
\end{eqnarray}
One should note that the energy levels of the chain system are also obtained as
\begin{eqnarray}\label{eq9}
E=\sum\limits_k {{\varepsilon }_k\left( \langle {\beta _k^\dag {\beta _k} \rangle - \frac{1}{2}} \right)}.
\end{eqnarray}

\begin{figure}
\centerline{\includegraphics[width=0.8\linewidth]{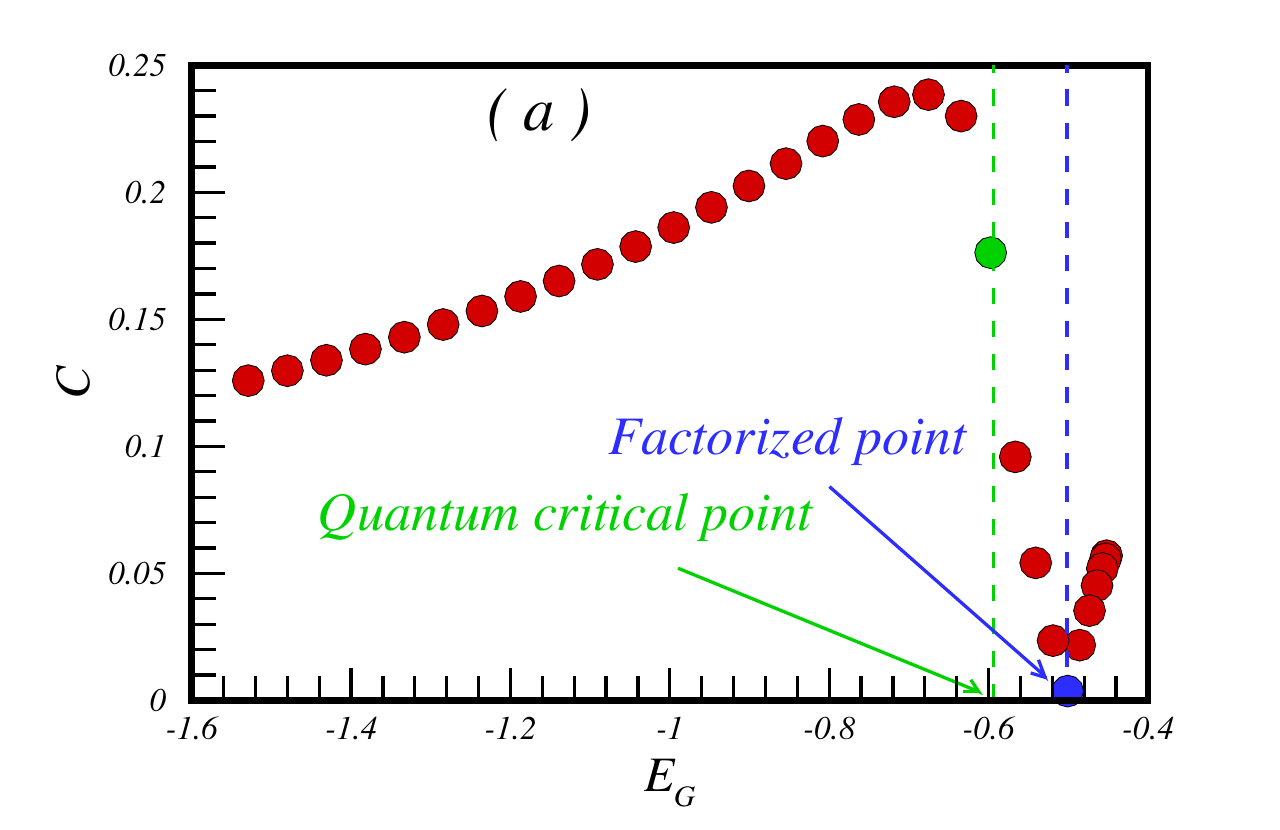}}
\centerline{\includegraphics[width=0.8\linewidth]{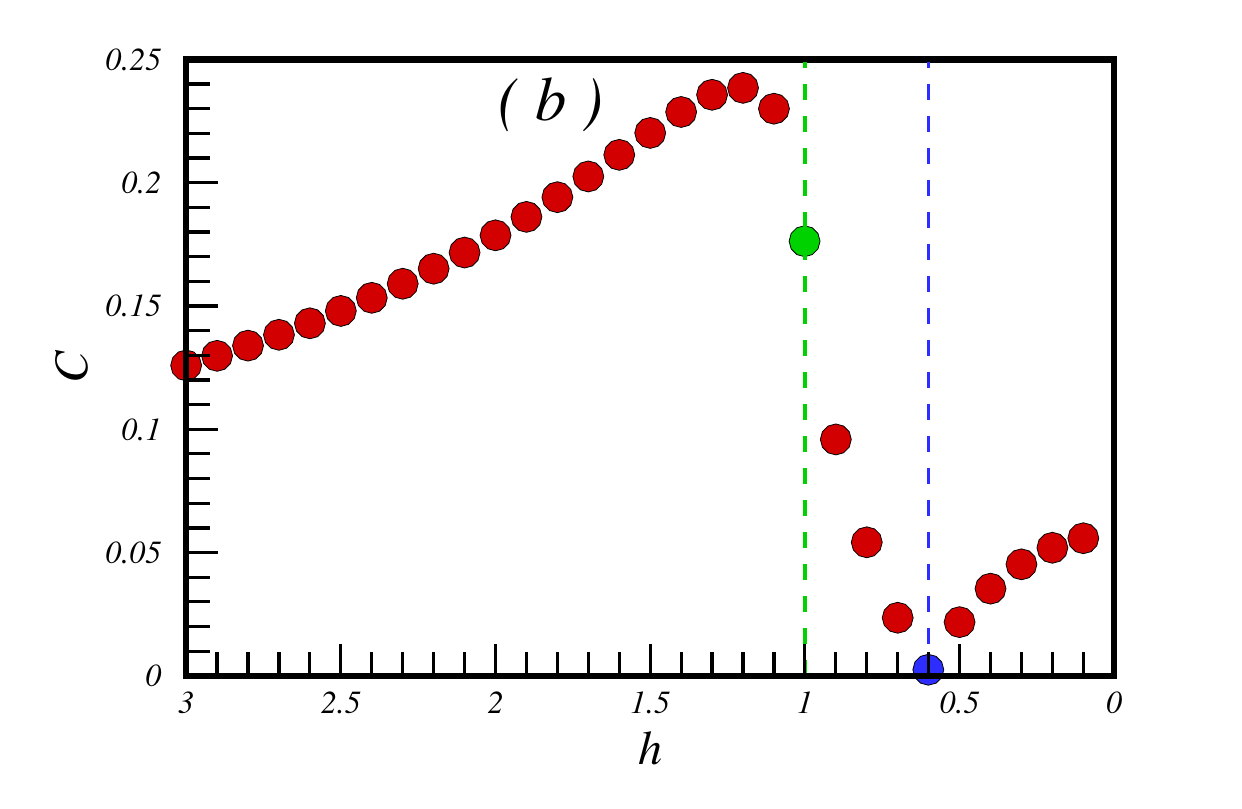}} 
\centerline{\includegraphics[width=0.8\linewidth]{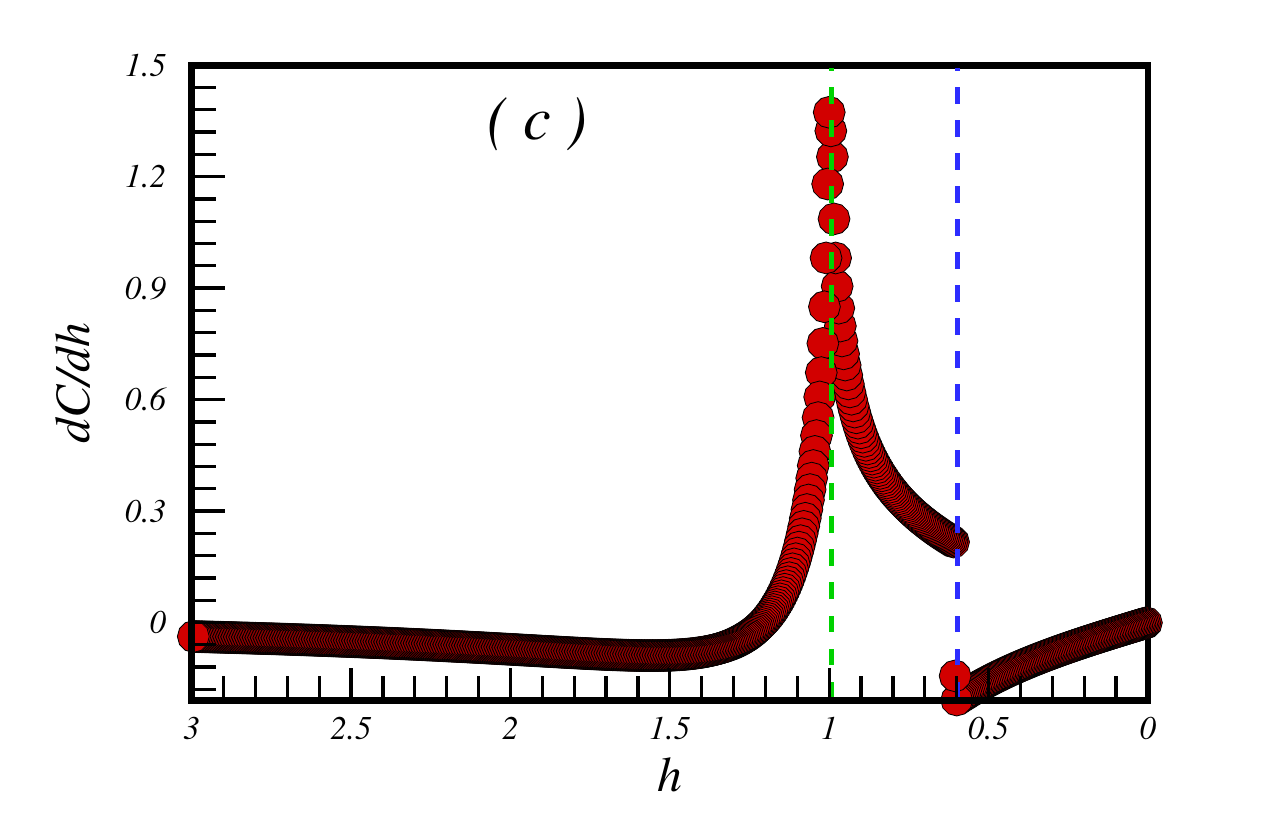}} 
\caption{(color online).   The concurrence between nearest-neighbour pair spins as a function of (a) the ground state energy and (b) the TF. It should be noted that the ground state exists in the subspace $N_B=0$ and in addition the concurrence disappears at the factorized point. (c) The first derivative of the concurrence with respect to the TF.  } 
\label{Fig1}
\end{figure}
Before delving into our findings regarding the correlation of spins in excited states, it's essential to examine the impact of the TF on the ground state of the model. For this investigation, we utilize a chain system with $N = 1000$ and $\delta=0.8$, varying the TF from $0$ to $3.0$.  The same calculations were performed for chains with both odd and even numbers of spins. We discovered that the concurrence remains independent of the parity of the number of spins, a result expected for very large systems.

Fig.~\ref{Fig1} (a) and (b) show how the results depend on the ground state energy and the TF. It is important to note that the ground state energy decreases with increasing TF, and this trend holds significance for our analysis. We observe that the model's ground state is entangled throughout the parameter space, except for a special point. At this particular point, the system undergoes factorization, occurring at $h_f=\sqrt{1-\delta^2}=0.6$ (with $E_G\sim -0.5$) [{\color{blue}\onlinecite{E17, E18,E19}}]. The concurrence increases as the field surpasses the factorized point, but it does not reach a maximum at the quantum critical field $h_c=1.0$. It is well-known that quantum correlation measures do not necessarily attain a maximum precisely at the quantum critical point. Instead, they often exhibit non-analytic behavior at criticality [{\color{blue}\onlinecite{E1, E17, E19-4, E19-5}}]. It implies that quantum correlation measures are not smooth functions of the control parameter and may exhibit discontinuities or singularities at the quantum critical point.  This characteristic is evident when plotting the first derivative of the concurrence with respect to the TF, as illustrated in Fig.~\ref{Fig1} (c). The derivative exhibits a sharp peak at the quantum critical point, signifying non-analytic behavior in the concurrence.

\begin{figure}
\centerline{\includegraphics[width=0.75\linewidth]{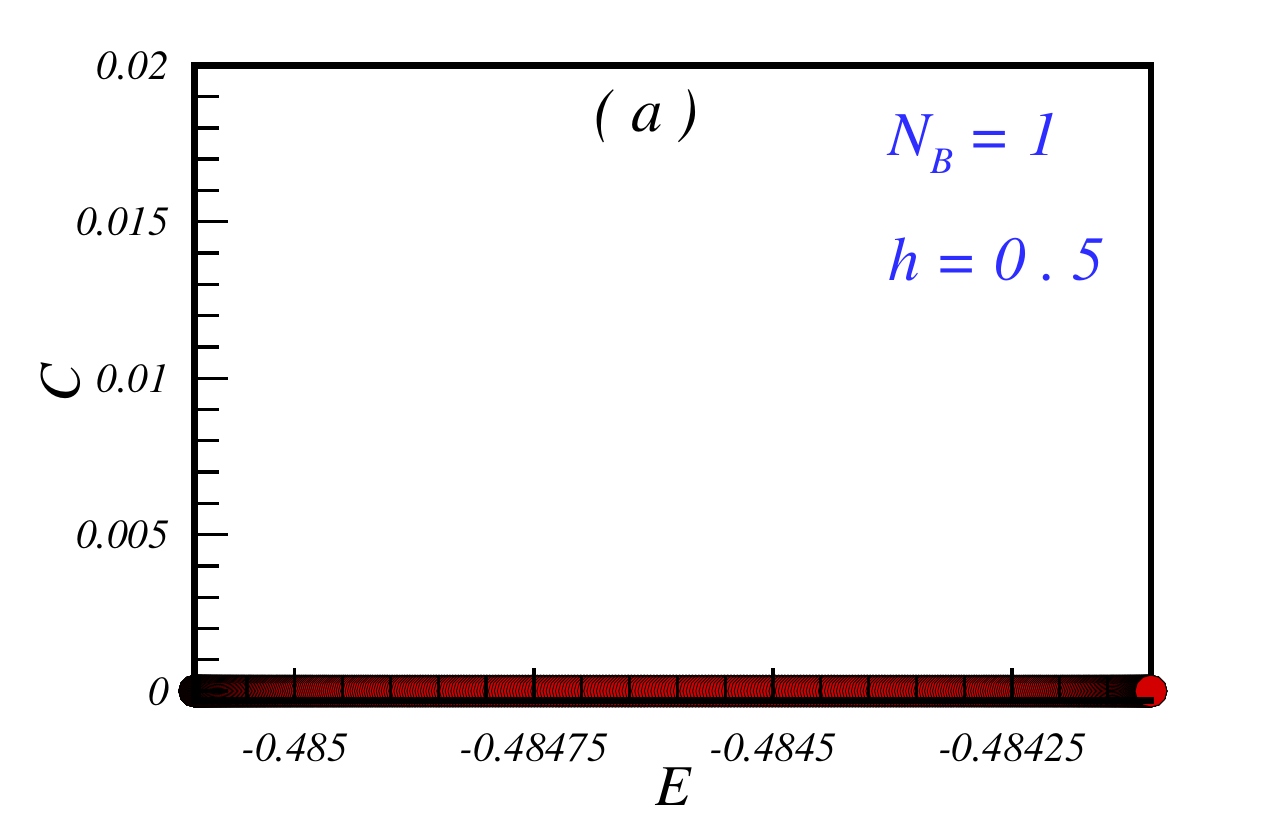}}
\centerline{\includegraphics[width=0.75\linewidth]{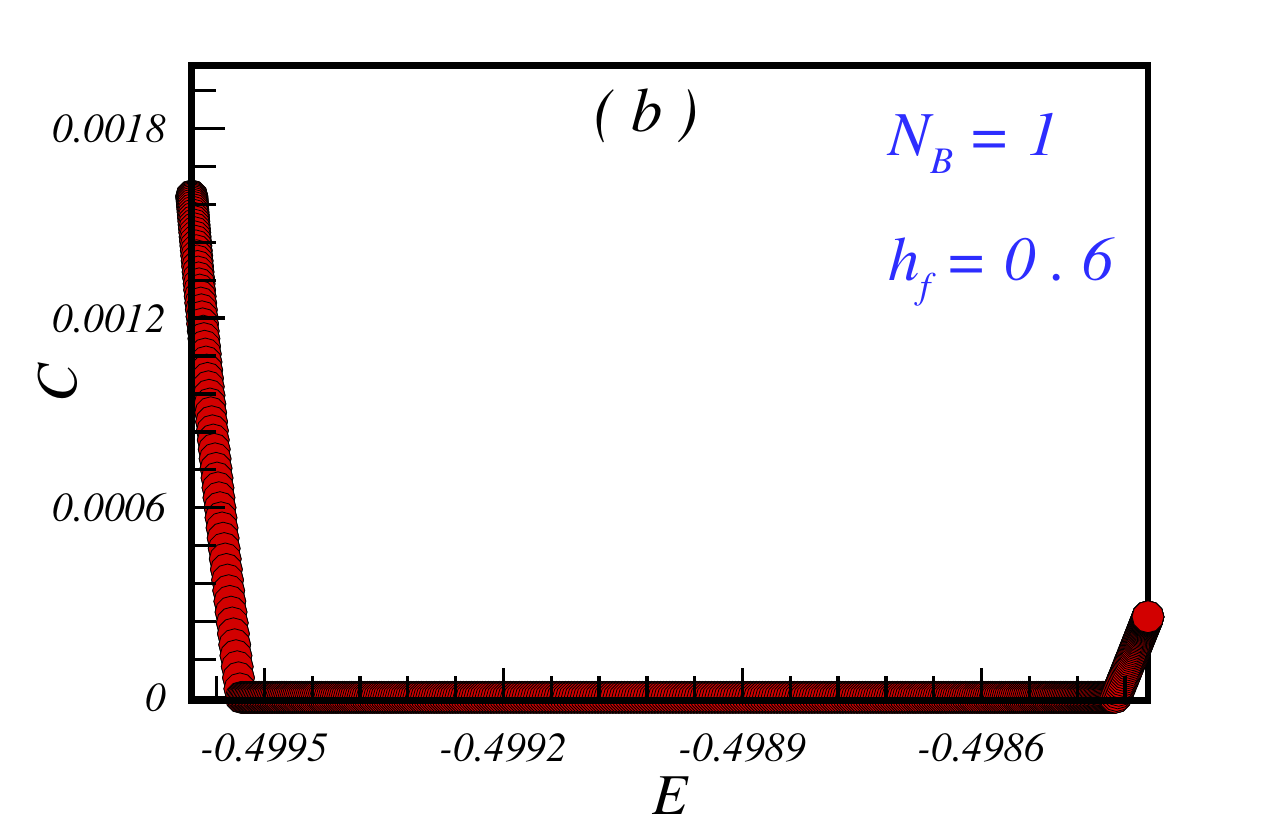}} 
\centerline{\includegraphics[width=0.75\linewidth]{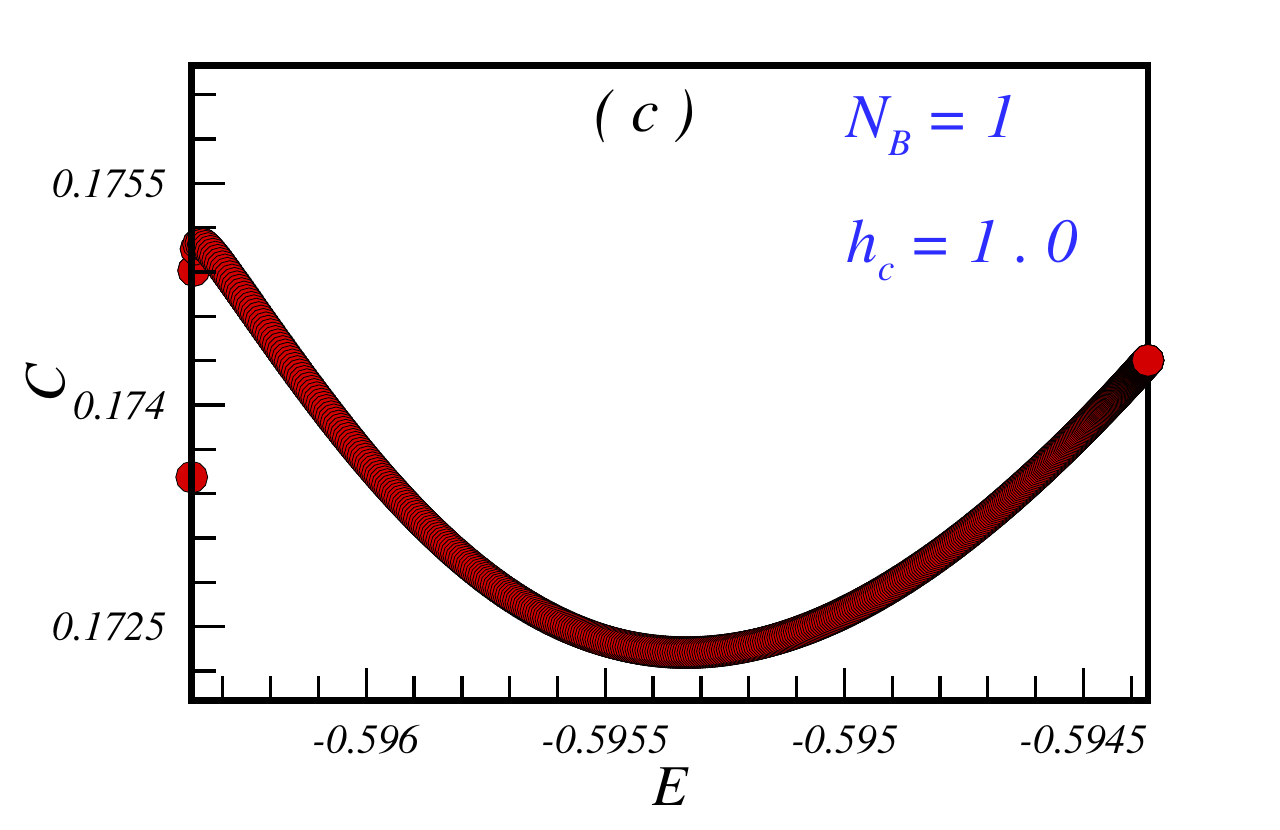}}
\centerline{\includegraphics[width=0.75\linewidth]{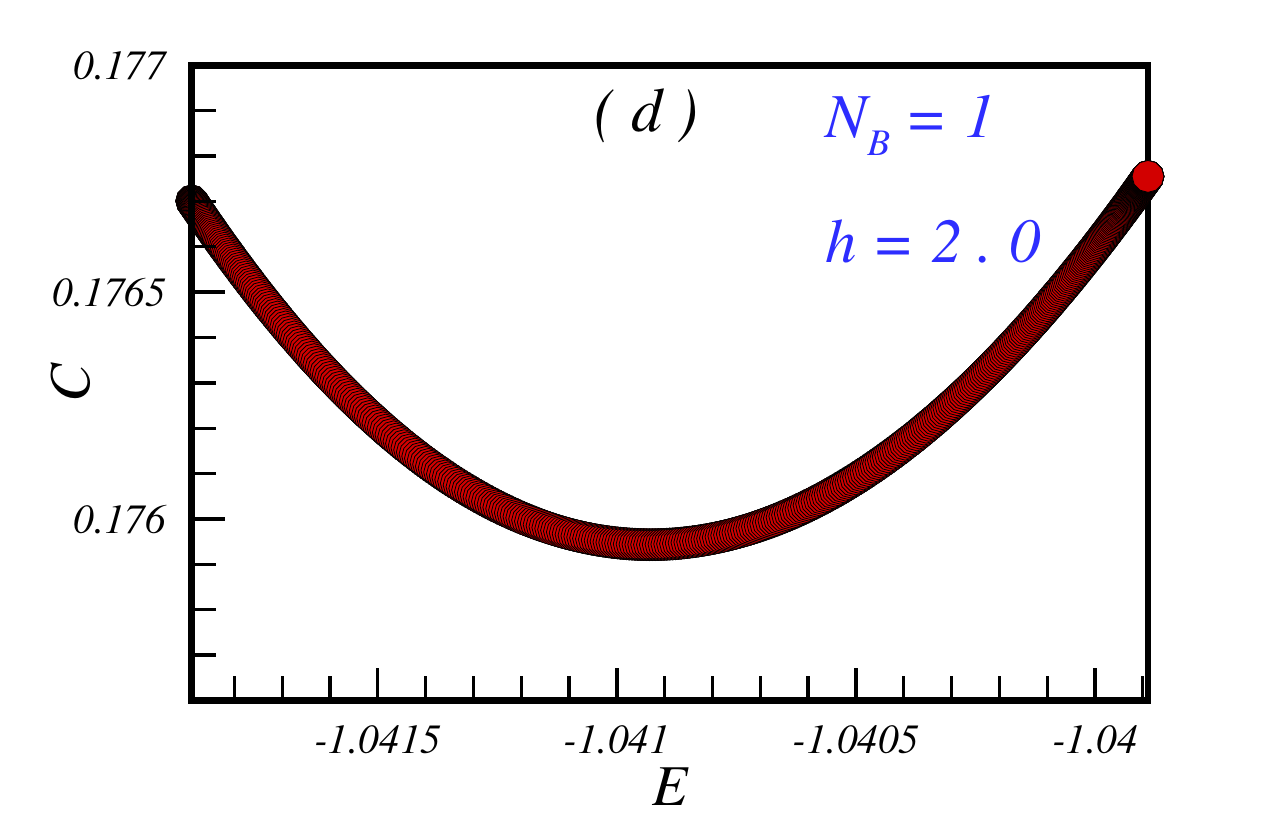}}
\includegraphics[width=0.75\linewidth]{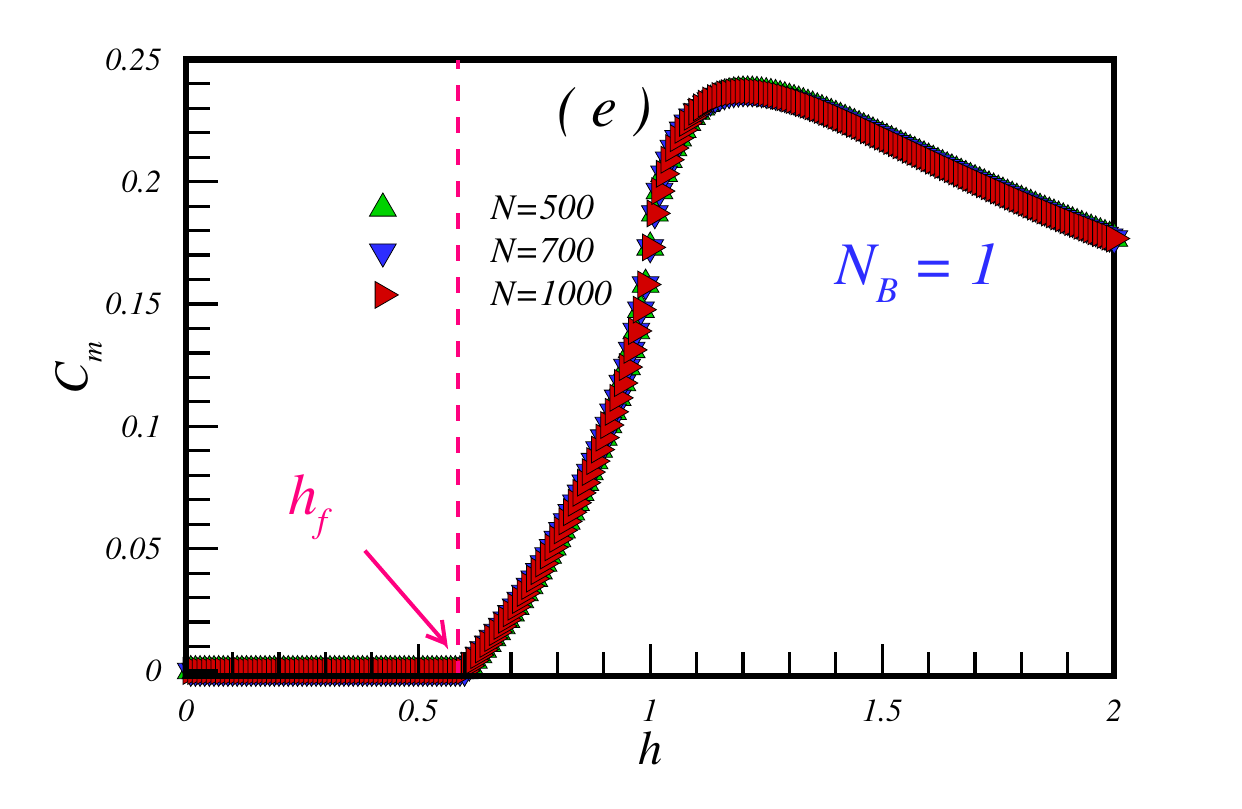}
\caption{(color online). The concurrence between nearest-neighbour pair spins as a function of the energy in the subspace with one Bogoliubov fermion, $N_B=1$ and chain size $N=1000$. (a) TF $h=0.5$ below the factorized point, (b) TF equal to factorized $h_f=0.6$, (c) TF at the quantum critical point $h_c=1$ and above the factorized point, (d) TF $h_c=2$ much higher than the factorized point. (e) The maximum concurrence in the subspace with one Bogoliubov fermion, $N_B=1$ with respect to the TF and different chain sizes $N=500, 700, 1000$. 
 }
\label{Fig2}
\end{figure}

Next, we explore the variations in concurrence when the system is in an excited state. The results are depicted in Fig.~\ref{Fig2} for a chain size $N=1000$ and a subspace with $N_B=1$. It is important to note that we conducted additional calculations for the concurrence in subspaces with up to $N_B=5$ fermions, and consistent behavior was observed across all subspaces. As Fig.~\ref{Fig2} (a) shows, there is no concurrence between nearest-neighbour spins in any excited state when $h<h_f$. Using $h=0.5$ as an example. Weak concurrence is predominantly  concentrated in very low-energy states at the factorized field, as illustrated in Fig.~\ref{Fig2} (b). Notably, there is no  nearest-neighbor pairwise entanglement in the middle states in this subspace. As the TF increases, the concurrence strengthens, leading to entanglement in all excited states, as evident in Fig.~\ref{Fig2} (c-d). The maximum concurrence for the subspace with $N_B=1$ is illustrated as a function of the TF in Fig.~\ref{Fig2} (e). We employed various chain sizes ($N=500, 700, 1000$) and observed no size effect. Nearest-neighbor spin pairs remain unentangled in all chain systems until the factorized field $h_f$ is reached. Additionally, we noted that the concurrence of excited states increases with the anisotropy parameter $\delta$, while the behavior remains consistent across different values of $\delta$.

In our pursuit to better comprehend the nature of the excited states in the model, we have introduced the concept of the density of entangled states in a subspace $N_B=m$. To achieve this, we initially calculate the width of the spectrum in the subspace $N_B=m$ as 
$\Delta=E_{max}-E_{min}$. Subsequently, we partition this width into $m'$ equal parts, denoted as $\Delta'=\frac{\Delta}{m'}$.  Finally, we define the density of the entangled states as

\begin{eqnarray}\label{eq2}
DoES(E)=N_C~\frac{m! (N-m)!}{N!},
\end{eqnarray}
where $N_C$ is the number of entangled states in the interval $E+\Delta'$. We also look at the distribution of the concurrence. 
To achieve this we partition the width of the entanglement range into $m"$ equal parts and then count the number of pairwise entangled states, $N'_{C}$, within  each part. The distribution of the concurrence is represented by 
\begin{eqnarray}\label{eq2}
Dis=N'_{C}~\frac{m!(N-m)!}{N!}.
\end{eqnarray}  

\begin{figure}
\centerline{\includegraphics[width=0.7\linewidth]{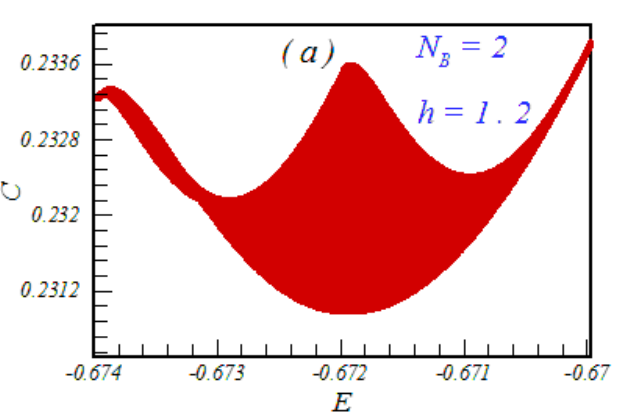}}  
\centerline{\includegraphics[width=0.8\linewidth]{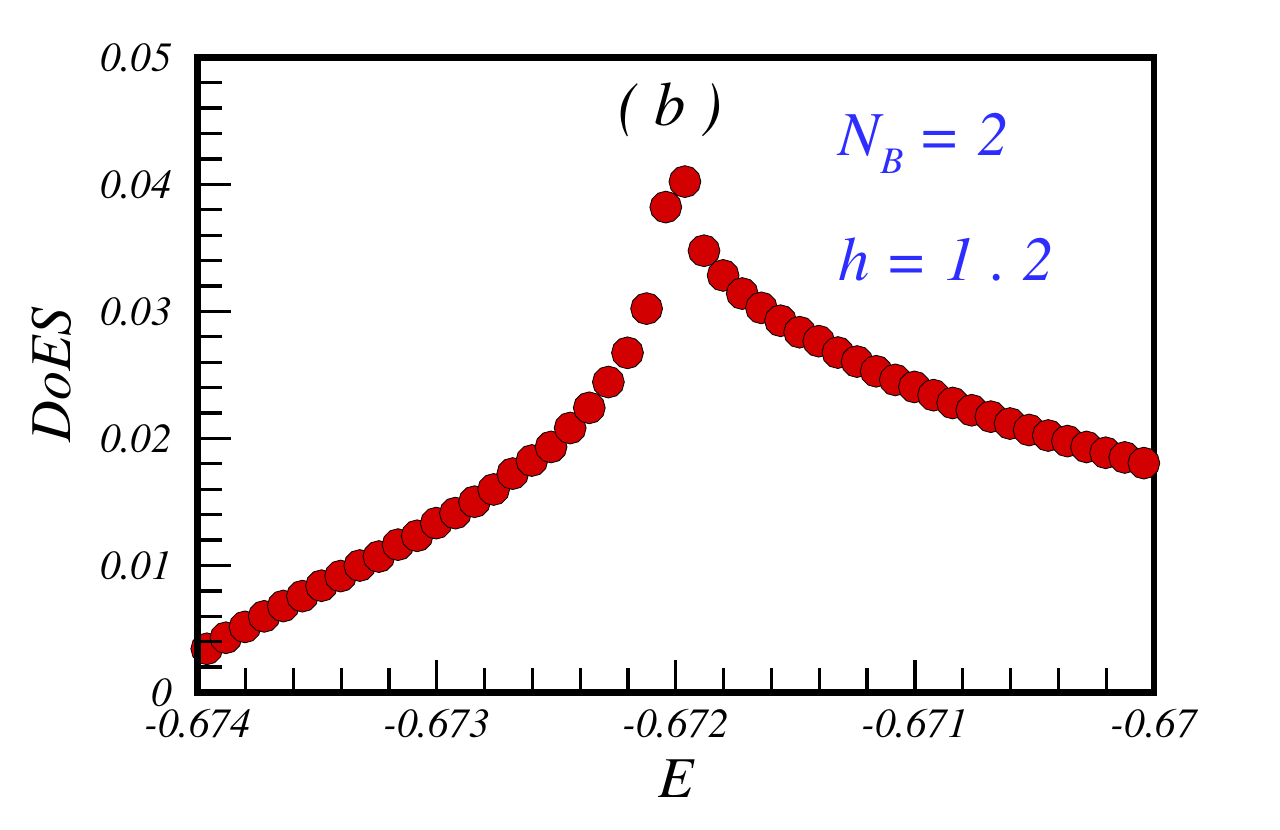}}
\centerline{\includegraphics[width=0.8\linewidth]{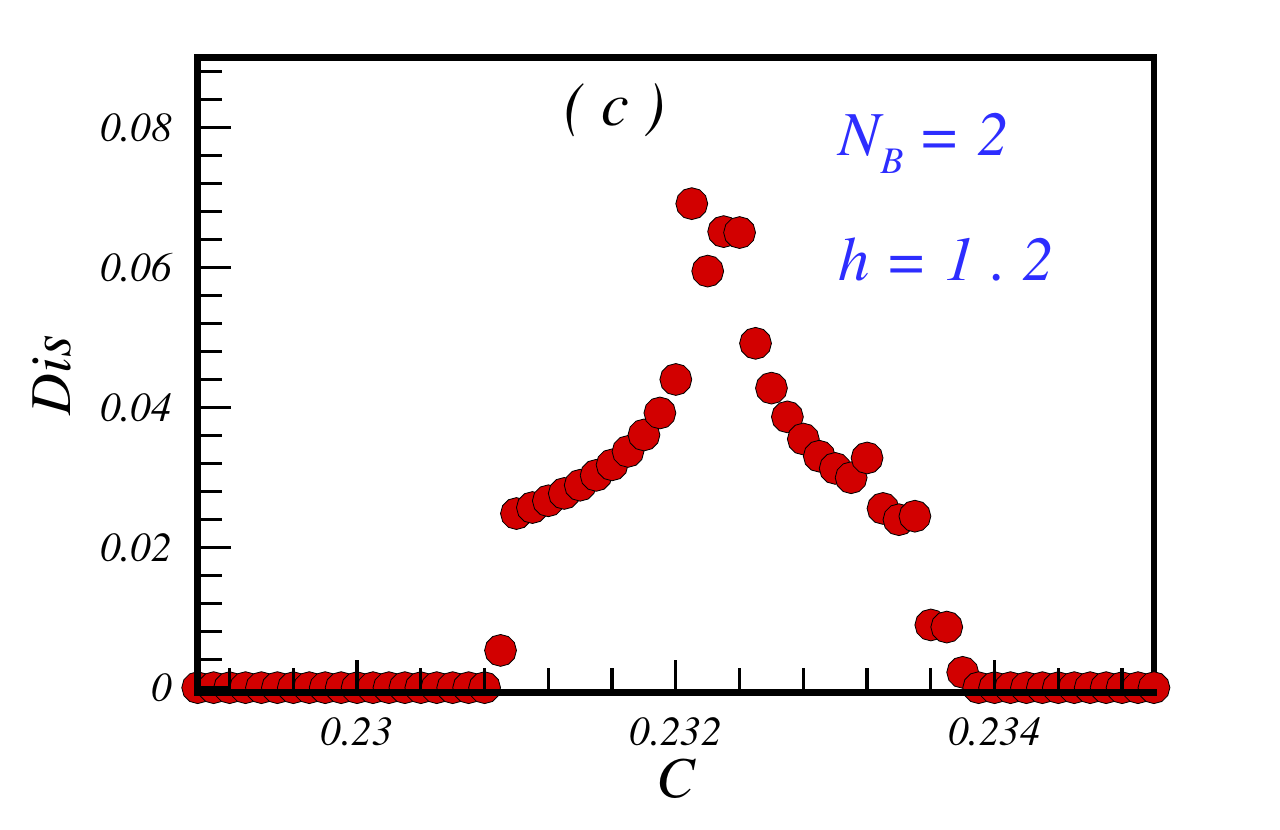}}
\caption{(color online). (a) The concurrence between nearest-neighbour pair spins as a function of the energy in the subspace with two Bogoliubov fermion, $N_B=2$ and chain size $N=1000$. Transverse field is $h=1.2$ higher than the factorized and quantum critical points. (b) The density of the entangled states as a function of the excited energies in the same subspace. (c) The distribution of the concurrence on its spectrum in the same subspace. }
\label{Fig3}
\end{figure}

The results are graphically illustrated in Fig. 3 for a constant transverse field $h=1.2$ and a chain size $N=1000$. We focus on calculating the excited states within the subspace $N_B=2$. As previously mentioned, all the excited states in each subspace exhibit entanglement, as depicted in Fig.~\ref{Fig3} (a) specifically for the subspace $N_B=2$.  The concurrence range is notably narrow ($\simeq 0.2355-0.2325=0.0030$), and numerous excited states with high concurrence span the entire energy spectrum. This observation aligns with the characteristics of integrable systems, quantum systems solvable precisely through analytical methods due to their abundant symmetries or conserved quantities [{\color{blue}\onlinecite{E20}}].

In Fig.~\ref{Fig3} (b), we present the density of entangled states (chosen as $m'=50$), reaching its peak at the center of the excited spectrum. Notably, as we shift the energy away from the spectrum's midpoint, the density of entangled states drops more rapidly in the lower excited states compared to the higher excited states. To conclude our investigation, Fig.~\ref{Fig3} (c) illustrates the distribution of concurrence (chosen as $m"=10000$). Notably, the most entangled excited states are prominently concentrated in the middle range where the concurrence is non-zero.

We can interpret our findings through the lens of the concept of typicality, asserting that the majority of states in a large Hilbert space share similar properties such as energy, entropy, and entanglement[{\color{blue}\onlinecite{E20-1,E20-2}}]. These 'typical states' represent the prevalent states within the Hilbert space and often have energies close to the average, typically situated at the center of the energy spectrum. To assess the density of entangled states in the energy spectrum, we examine the nearest-neighbor pairwise entanglement of typical states. Our results consistently reveal that typical states exhibit significant entanglement between nearest-neighbor pairs of spins and, in general, are highly entangled. Consequently, the density of entangled states is notably high at the center of the energy spectrum. This pattern arises because typical states, being complex and random, embody substantial information in their correlations. Importantly, this outcome holds true for generic quantum systems lacking special symmetries or constraints that might otherwise reduce their complexity.

 Alternatively, our results may find partial explanation in the context of the eigenstate thermalization hypothesis [{\color{blue}\onlinecite{E21,E22,E23}}]. This hypothesis provides insights into how an isolated quantum system can attain a state of thermal equilibrium even without interacting with its environment. According to this hypothesis, the matrix elements of any observable in the energy eigenbasis exhibit a specific structure resembling random numbers. Consequently, the expectation value of any observable in an energy eigenstate closely approximates the thermal average, with minimal fluctuations. As depicted in Fig.~\ref{Fig3} (a), our findings align with the strong eigenstate thermalization hypothesis, indicating that for a sufficiently large chain system (e.g., $N=1000$), the concurrence between nearest-neighbor pair spins in subspace $N_B=2$ smoothly depends on the energy, rather than being contingent on the specific eigenstate chosen within this subspace.

\section{conclusion}

We conducted a study on the one-dimensional spin-1/2 XY model incorporating a TF. To diagonalize the Hamiltonian, we employed the fermionization technique. Notably, the diagonalized Hamiltonian and the total Bogoliubov number operator share identical eigenstates due to their commutative nature. The vacuum state of the Bogoliubov number operator corresponds to the ground state of the chain system, while the excited states are confined to subspaces with fixed values.

This model exhibits a well-established ground state with two distinct phases. For $h<h_c=J$, there exists an antiferromagnetic phase characterized by order, while for $h>h_c$, a paramagnetic phase with disorder emerges. Quantum fluctuations prevent saturation at zero temperature. Additionally, there is a noteworthy point, $h_{f}^{2}+(J\delta)^2 = J^2$, termed the factorized point. At this point, the ground state becomes one of the eigenstates of the total spin’s $z$-component. This ground state at the factorized point lacks quantum correlation, such as entanglement, and is separable from a quantum information perspective. However, excluding the factorized point, entanglement persists among nearest-neighbor spin pairs throughout the entire ground state phase diagram.

Recent studies have highlighted the significance of the factorized point as a boundary in this model, particularly concerning two distinct regions. The first region pertains to the Loschmidt echo, suggesting the possibility of complete revivals of quantum states after a quantum quench within a time period proportional to the system size [{\color{blue}\onlinecite{E24}}]. However, it has been demonstrated that full revivals are absent in the one-dimensional spin-1/2 XY model with a TF [{\color{blue}\onlinecite{E19-1}}]. Instead, two distinct regimes with different behaviors were identified. The quasi-particle picture with maximum group velocity applies well for $h>h_f$, but not for $h \leq h_f$. In this latter region, the revivals cannot be explained by quasi-particles with fixed velocities. Another aspect related to the factorized point is the spin squeezing parameter. At zero temperature, the system's behavior is dictated by its ground state, revealing two distinct regimes of squeezing in the ground state phase diagram:  $h <h_f$ with no spin squeezing, and $h >h_f$ with spin squeezing.

In this study, we have examined the distribution of concurrence in the excited states of the chain model. Specifically, we computed the concurrence between nearest-neighbor spins in all excited states above the ground state. For $h<h_f$, no nearest-neighbor pairwise entanglement is observed in any excited state. At the factorized point, we identify weak concurrence between nearest-neighbor spins in very low energy states. Conversely, for $h > h_f$, the concurrence increases, and all excited states exhibit entanglement. Notably, the entangled states are most densely concentrated at the center of the excited spectrum.

We have also provided an explanation for our results based on the eigenstate thermalization hypothesis. According to this hypothesis, matrix elements of any observable in the energy eigenbasis exhibit a structure reminiscent of randomness. Consequently, the expectation value of any observable in an energy eigenstate closely approximates the thermal average, with minimal fluctuations. Our findings align with this hypothesis, indicating that the concurrence between nearest-neighbor spins varies smoothly with the energy in a subspace, and importantly, it does not rely on the specific choice of an excited eigenstate.


\end{document}